\documentclass[aps,prl,reprint,groupedaddress]{revtex4-2}
\usepackage{amsmath}
\usepackage{amsthm}
\usepackage{mathtools}
\usepackage{graphicx}
\usepackage{amsfonts}
\usepackage{amssymb}

\newcommand{\transp}{\mathsf{T}}
\renewcommand{\vec}[1]{\mathbf{#1}}

\DeclareMathOperator*{\argmax}{arg\,max}
\begin{document}
\title{Generalized Maximum Entropy Methods as Limits of the Average Spectrum Method}

\author{Khaldoon Ghanem}
\affiliation{Quantinuum, Leopoldstrasse 180, 80804 Munich, Germany}
\author{Erik Koch} 
\affiliation{J\"ulich Supercomputer Centre, Forschungszentrum J\"ulich, 52425 J\"ulich, Germany}
\affiliation{JARA High-Performance Computing, 52425 J\"ulich, Germany}

\date{\today}

\begin{abstract}
    
We show that in the continuum limit, the average spectrum method (ASM) is equivalent to maximizing Rényi entropies 
of order $\eta$, of which Shannon entropy is the special case $\eta=1$. 
The order of Rényi entropy is determined by the way the spectra are sampled.
Our derivation also suggests a modification of Rényi entropy, giving it a non-trivial $\eta\to0$ limit. We show that the sharper peaks generally obtained in ASM are associated with entropies of order $\eta<1$. Our work provides a generalization of the maximum entropy method
that enables extracting more structure than the traditional method.
\end{abstract}
	
\maketitle

Spectral reconstruction is an inverse problem where the goal is estimating a non-negative spectrum given incomplete and/or noisy data. 
It is a reccurring problem in many branches of physics and has many important applications including, but not limited to, image reconstruction, X-ray diffraction, mass spectrometry  and analytic continuation of quantum Monte Carlo data~\cite{Gull78, Wilkins84, Zahng97, Silver90}. Mathematically, the problem can be formulated as a Fredholm integral equation of first kind
\begin{equation}\label{eq:fredholm}
	g(y_j) = \int dx\ K(y_j, x) f(x)\;,
\end{equation}
where the left-hand side $g(y_j)$ represents $m$ data points known numerically, while the integral kernel $K(y_j, x)$ is a continuous function known analytically and determined by the problem at hand. The spectrum $f(x)$ is a non-negative and integrable  function $f(x)$ that needs to be estimated.
In this letter, we focus on problems where the norm of the spectrum is known apriori, so we can always rescale the above equation such that the spectrum is normalized to one.

The most widely-used approach for solving this problem is the maximum entropy method (MaxEnt)~\cite{Gull78, Silver90,Jarrell12}.
In its most basic form, it aims to find the spectrum maximizing Shannon entropy under known constraints 
\cite{Ghanem23}
\begin{equation*}
	f_\text{MaxEnt} = \argmax_{f \in \mathcal{C}} S[f],
\end{equation*}
where 
\begin{equation}\label{eq:shannon}
	S[f] \coloneqq  - \int dx \ f(x)  \ln \left(\frac{f(x)}{D(x)}\right)\;
\end{equation}
is Shannon entropy and represents the expected amount of information in a spectrum $f$ relative to another $D$, called the default model.
The manifold $\mathcal{C}$ represents the non-negativity and normalization constraints as well as the $m$ data constraints of \eqref{eq:fredholm}. In case of noisy data, the later is usually imposed via the discrepancy principle where the fit to the data is set equal to the expected level of noise~\cite{Morozov1984, Tikhonov77, Gull84, Ghanem23}.
Assuming Gaussain noise with zero mean and covariance matrix $\vec{L}$ and denoting the data residuals as $r_i \coloneqq g(y_i) - \int dx\ K(y_i, x) f(x)$, the fit is defined as $\chi^2[f] \coloneqq \vec{r}^\transp\vec{L}^{-1}\vec{r}$, and the constrained manifold reads
\begin{equation*}
    \mathcal{C} \coloneqq \left\{f(x)\geq 0: \int dx f(x) = 1,\; \chi^2(f)  = m \right\}\;.
\end{equation*}
MaxEnt can be implemented efficiently~\cite{Bryan90} and produces good results in general, which contributed to its widespread adoption.
Nevertheless, a common criticism, especially within the analytic continuation community, is that it produces spectra with overly smooth peaks.
In addition, one needs to pick a default model whose choice may bias the result.
Consequently, the average spectrum method (ASM) emerged as a promising alternative that addresses these issues~\cite{White91, Sandvik98, Sandvik16}.
It averages equally over all normalized non-negative spectra that fit the data constraints~\footnote{For noisy data, ASM  usually weights the spectra with how well they fit the data. Restricting the averaging to a specific data fit makes the comparison with constrained MaxEnt more seamless, but 
  it is straightforward to extend our arguments to that case.}
\begin{equation*}
	f_\text{ASM}(x) = \int_{\mathcal{C}} \mathcal{D}f \ f(x)\;.
\end{equation*}
The averaging was assumed to wash out features not supported by the data and thus produce unbiased results without the need for a default model.
It was shown later, however, that the above functional integral is not uniquely defined and that a default model is implied by the choice of the discretization grid~\cite{Ghanem20a}. Even a sampling of the grid points amounts to a change of the functional measure and thus produces a different result~\cite{Ghanem20b}.
Nevertheless, ASM remains a viable alternative to MaxEnt due to its apparent ability of producing sharper spectra~\cite{Sandvik98, Sandvik16, Gunnarsson07, Syljuasen08,Fuchs10, Ghanem17}.

A connection between MaxEnt and ASM was first suggested by Ref.~\cite{Beach04} which argued that the smoother spectra of MaxEnt are the result of it being a mean-field approximation of ASM. However, it was shown in Ref.~\cite{Ghanem20a} that such an approximation implies an entropy different from Shannon entropy used by MaxEnt. The new entropy has the same functional form of Shannon, but with the role of the default model and the spectrum reversed
\begin{equation}\label{eq:gk}
	S_{\operatorname{GK}}[f] \coloneqq  - \int dx \ D(x)  \ln \left(\frac{D(x)}{f(x)}\right)\;,
\end{equation}
where, following Ref~\cite{Shao23}, the subscript $\operatorname{GK}$ is used to distguish this specific form of entropy.
GK entropy was derived for the case of sampling spectral weights on a fixed grid, and Ref.~\cite{Shao23} recognized that sampling grid positions as well as spectral weights should lead to a yet different entropy and conjectured it to be a linear mixture of Shannon and GK entopies. 
One major contribution of our work here is deriving the correct entropic form for this case and generalizing it. The result turns out to be the family of Rényi entropies.

In his seminal work~\cite{Renyi61}, Rényi relaxed the axioms of Shannon and derived the entropies
\begin{equation}\label{eq:renyi}
S_{\operatorname{R}}[f, \eta] \coloneqq \frac{1}{1-\eta} \ln  \int dx \ \frac{f(x)^\eta}{D(x)^{\eta-1}} \;,
\end{equation}
where the parameter $\eta$ is called the order of the entropy. In the limit $\eta \to 1$, one recovers Shannon entropy as a special case. 
Rényi entropy was originally proposed as an alternative information measure and has since found important applications in various fields including statistical mechanics~\cite{Lenzi2000} and quantum information~\cite{Mueller13}.
Other generalized entropies have since been developed~\cite{Amigo18}, the most prominent being Tsallis entropy~\cite{Tsallis88} 
\begin{equation}
	S_{\operatorname{T}}[f, \eta] \coloneqq  \frac{1}{\eta-1} \left[\int dx \ \frac{f(x)^\eta}{D(x)^{\eta-1}} - 1 \right]   \;.
\end{equation}
Rényi and Tsallis entropies are related by a monotonic mapping, so maximizing one is equivalent to maximizing the other~\cite{Wong22}.
The existence and usefulness of these entropies 
calls into question the uniqueness of Shannon entropy within MaxEnt, which has been the subject of extensive debate~\cite{Shore80, Skilling88,Karbelkar86, Uffink95, Burnier13, Jizba19}.
The results in this letter supplement these arguments and give a systematic approach for deriving generalized entropies as limits of ASM\@.
We also show how a simple rescaling of Rényi entropy changes the limit $\eta\to0$ from simply vanishing to giving the GK entropy.
Finally, we discuss the effect of maximizing generalized entropies of different orders, dubbed as MaxGEnt, and examplify it in an illustrative test case.

\paragraph{Limits of ASM:} To evaluate the functional integral of ASM, we discretize the spectrum by representing it as a linear combination of $N$ delta functions
\begin{equation}
	f(x) = \sum_{i=1}^N f_i \delta(x-x_i)\;,
\end{equation}
where $x_i$ are their positions and $f_i$ are their non-negative weights that sum up to one. 
The average spectrum is obtained by sampling the weights and/or positions, binning each spectrum on a discretized grid, and averaging the binned spectra \cite{Beach04,Shao23}.
Let the binning grid have $M$ intervals denoted as $\mathcal{I}_j$, then the amplitude of a binned sample on the $j$-th bin is the of integral  of the sample over that bin  $F_j \coloneqq \int_{\mathcal{I}_j} dx f(x) = \sum_{x_i \in \mathcal{I}_j} f_i$.
Clearly, different samples can give rise to the same binned amplitudes.
In physical terms, the unbinned sample $(x_1, \dots, x_N; f_1, \dots, f_N)$ can be thought of as a microstate, while the binned one $(F_1, \dots, F_M)$ as the macrostate.
The probability distribution of a macrostate is completely determined by how the microstates are sampled.
As the number of delta functions increases, this distribution becomes singular but its normalized logarithm has a well-defined limit 
\begin{equation*}
	\lim_{N\to\infty} \frac{\ln P\left(F_1, \ldots, F_M\right)}{N} =: 
 H\left(F_1, \ldots, F_M\right),
\end{equation*}
which is defined as the entropy associated with that specific way of sampling~\cite{Shao23}.
In that limit, averaging binned spectra becomes equivalent to finding the spectrum with the maximum entropy.
When the weights are sampled uniformly while the positions $x_i$ are fixed and placed with a density $D(x)$, one obtains GK-entropy~\cite{Ghanem20a}.
On the other hand, when the weights are held fixed at $1/N$ while positions are sampled independently from a default model $D(x)$, it yields Shannon entropy~\cite{Beach04,Shao23}.
In the following, we seek the entropy of sampling both the positions and weights simultaneously.
Let $n_j$ be the number of delta functions falling in the $j$-th bin, then the desired probability can be decomposed as
\begin{equation}\label{eq:total_prob}
		P\left(\left\{ F_j\right\}\right) = \sum_{\left\{ n_j\right\}} P\left(\left\{ n_j\right\}\right)  \times 
		P\left(\left\{ F_j\right\}| \left\{ n_j\right\}\right)\;,
\end{equation}
where $P\left(\left\{ n_j\right\}\right) $ is the probability that a configuration of $N$ delta functions is distributed as $n_1, \ldots, n_M$ over the bins $\mathcal{I}_1, \ldots, \mathcal{I}_M$, and $P\left(\left\{ F_j\right\}| \left\{ n_j\right\}\right)$ is the conditional probability that the weights of this configuration of delta functions adding up to $F_1, \ldots, F_M$ for the respective bins.

\paragraph{Configuration Probability:}
The positions are sampled independently from $D(x)$.
The probability of a delta function falling inside the $j$-th bin  is $D_j \coloneqq \int_{\mathcal{I}_j} dx D(x)$, so the probability of having a specific configuration $\left\{ n_j\right\}$ follows the multinomial distribution
\begin{equation*}
	P\left(\left\{ n_j\right\}\right) = \frac{N!}{n_1! \ldots n_M!}\  D_1^{n_1} \ldots D_M^{n_M}\;.
\end{equation*}
We then use Stirling's formula $\ln x! \approx x \ln x - x$ to get the following approximation to the configuration probability
\begin{equation}\label{eq:config_prob}
	P\left(\left\{ n_j\right\}\right)  \approx 
    \exp \left[- \sum_j n_j\ln \frac{n_j}{N D_j}  \right]
    \;.
\end{equation}

\paragraph{Conditional  Probability of Amplitudes:} 
The weights $f_i$ are sampled under the condition of being non-negative and summing up to one. Therefore, they follow a Dirichlet distribution with unit shape parameters \cite{Ghanem20a}.
According to the aggregation property of the Dirichlet distribution, the partial sums of random variables following the Dirichlet distribution also follow a Dirichlet distribution where the corresponding shape parameters are summed~\cite{dirichlet}.
Assuming $n_j$ delta function fall in the $j$-th bin, the shape parameter of amplitude $F_j$ is $\sum_{x_i \in \mathcal{I}_j} 1 = n_j$ and the desired Dirichlet distribution reads
\begin{equation*}
	P\left(\left\{ F_j\right\}| \left\{ n_j\right\}\right)= \frac{\Gamma( N)}{\Gamma( n_1) \ldots \Gamma( n_M)}  \ F_1^{ n_1-1} \ldots F_M^{ n_M-1}\;,
\end{equation*}
where $\Gamma(z)$ is the gamma function. In the limit of very large $N$, we can replace $ n_i$ by $ n_i{+}1$ and use Stirling's formula again to get an approximation to the conditional probability
\begin{equation} \label{eq:cond_prob}
	P\left(\left\{ F_j\right\}| \left\{ n_j\right\}\right) \approx \exp \left[-  \sum_j n_j \ln \frac{n_j}{N F_j}\right]\;.
\end{equation}

\paragraph{Total Probability of Amplitudes:} The total probability in 
Eq.~\eqref{eq:total_prob} is a sum over the product of Eqs.~\eqref{eq:config_prob} and \eqref{eq:cond_prob} for all possible configurations. In the limit $N\to \infty$, the sum is dominated by the term with the largest exponent.
Findig its value is simple when defining the auxiliary quantities  
\begin{equation*}
    E \coloneqq \sum_j \sqrt{F_j D_j}, \quad\mbox{and }\quad
    E_j \coloneqq \sqrt{F_j D_j}/E \;,
\end{equation*}
and writing the total probability in terms of normalized configurations $\tilde{n}_j \coloneqq n_j/N$ as
\begin{equation*}
	P\left(\left\{ F_j\right\}\right) = \exp \left[  2 N \ln E  \right] \!\!\sum_{\tilde{n}_1,\ldots, \tilde{n}_M}\!\!  \exp \left[   -2 N \sum_j  \tilde{n}_j \ln \frac{\tilde{n}_j}{E_j} \right].
\end{equation*}
We recognize the exponent above as a multiple of the Shannon entropy of a normalized distribution, $\tilde{n}_j$, with respect to another, $E_j$.
Its maximum value is zero and thus the whole sum of exponentials is dominated by one. In the limit $N\to\infty$, we can therefore write
\begin{equation}\label{eq:log_prob}
	P(F_1,\ldots, F_M) \approx \exp\left[2 N \ln E \right]\;,
\end{equation}
from which we find the entropy of sampling both weights and positions
\begin{equation*}
	H(F_1, \dots, F_M) = 2 \ln \left( \sum_j \sqrt{D_j F_j}\right)\;.
\end{equation*}
This is a discrete version of Rényi entropy~\eqref{eq:renyi} with $\eta=1/2$. Taking the limit of fine binning grid $M\to \infty$, one can also recover the corresponding continuous version.

\paragraph{Generalized Entropies:} 
It is rather interesting that sampling both weights and positions gives an entropy that is symmetric in both the spectrum and the default model, while sampling one of them gives the same entropic form but with the role of default model and spectrum reversed (compare Eq.~\eqref{eq:shannon} with \eqref{eq:gk}).
A natural question then is to ask whether other ways of sampling could give rise to a general form that connects the different entropies. 
This is indeed the case if we replace the uniform distribution of the weights with a symmetric Dirichlet distribution of arbitrary shape parameter $\alpha > 0$.
The shape parameter controls how concentrated the weights are around their mean value $1/N$.
The uniform distribution is the special case with $\alpha=1$, and in the limit $\alpha\to\infty$, the sampling becomes equivalent to ASM with fixed weights.
We can derive the entropy of a general shape parameter $\alpha$ in a similar way to the derivation above, which corresponds to $\alpha=1$.
In the general case, the configuration probability remains the same, while the  conditional probability of amplitudes gets raised to the power $\alpha$ so that the total probability reads 
\begin{equation*}
	\begin{split}
	P_\alpha\left(\left\{ F_j\right\}\right) = &\exp \left[  (\alpha+1) N \ln E^\prime  \right] \ \times \\
	&\sum_{\tilde{n}_1,\ldots, \tilde{n}_M}  \exp \left[   -(\alpha+1) N \sum_j  \tilde{n}_j \ln \frac{\tilde{n}_j}{E^\prime_j} \right]\;,
	\end{split}
\end{equation*}
where we defined the generalized auxiliary quantities
\begin{equation*}
   E' \coloneqq \sum_j (F_j^\alpha D_j)^{\frac{1}{1+\alpha}}
   \quad\mbox{and }\quad
   E'_j \coloneqq {(F_j^\alpha D_j)^{\frac{1}{1+\alpha}}}/{E'},\; 
\end{equation*}
Using a similar argument to the $\alpha=1$ case, we find the entropy in the general $\alpha$ case to be
\begin{equation*}
		H_\alpha\left(F_1, \ldots, F_M\right) = (\alpha+1) \ln \left[\sum_j F_j \left(\frac{F_j}{D_j}\right)^{\frac{-1}{1+\alpha}} \right]\;. 
\end{equation*}
which is the Rényi entropy of order $\eta = \alpha/(1{+}\alpha)$.

\paragraph{Relating GK entropy to Rényi entropy:}
In the limit of $\alpha\to \infty$, which gives fixed weights, the order of the entropy has the limit $\eta \to 1$ and we recover Shannon entropy as expected.
One would then expect to recover GK entropy in the other limit $\eta \to 0$.
Instead, Rényi entropy becomes trival in that limit and vanishes when $f$ has the same support as $D$.
In fact, this limit corresponds to $\alpha \to 0$ where the earlier use of Stirling's approximation for the conditional probability breaks down.
Nevertheless, we can still recover GK entropy by slightly modifying the denfintion of Rényi entropies as
\begin{equation*}
	S_{\operatorname{\tilde{R}}}[f, \eta] \coloneqq \frac{1}{\eta}  S_{\operatorname{R}}[f, \eta]  = \frac{1}{\eta(1-\eta)} \ln  \int dx \ \frac{f(x)^\eta}{D(x)^{\eta-1}} \;.
\end{equation*}
In the limit $\eta \to 0$, this modification gives GK entropy and thus can be seen as having a richer structure than the original expression~\eqref{eq:renyi}.
Crucially, it does not change the properties of Rényi entropy at any non-vanishing order, merely rescaling it.

\paragraph{Maximizing gerneralized entropies (MaxGEnt):}
It is well-known that maximizing Shannon entropy under linear constraints leads to distributions of exponential form, while maximizing Rényi (or equivalently Tsallis) entropies gives power-law distributions~\cite{Tsallis88, Curado91}.
For convenience, instead of maximizing the modified Rényi entropy itself, we maximize its monotonic transformation
\begin{equation}
S_{\operatorname{\tilde{T}}}[f, \eta] \coloneqq \frac{e^{\eta(1-\eta) S_{\operatorname{\tilde{R}}}[f, \eta]-1}}{\eta(1{-}\eta)}, 
\end{equation}
representing our modified version of Tsallis entropy with a non-trivial $\eta \to 0$ limit.
We then define the Lagrangian
\begin{equation}
	\begin{split}
	\mathcal{L}_\eta \left[f, \mu, \{\lambda_k\}\right] =  S_{\tilde{T}}[f, \eta] - \mu \left[\int dx f(x) -1\right]&\\
	  - \mu (\eta{-}1) \sum_k \lambda_k \left[\int dx f(x) c_k(x) - C_k\right]& \;,
	\end{split}
\end{equation}
where $\int dx f(x) c_k(x) = C_k$ are a set of linear constraints, and $\mu, \lambda_k$ are Lagrange multipliers. Note that the $\lambda_k$ are explicitly multiplied by $\mu (\eta-1)$ for convince.
Rényi/Tsallis maximizers are saddle points of this Lagrangian and have the general form
\begin{equation}
	f(x) = \frac{1}{Z}D(x) \left[1+(1{-}\eta)\sum_i \lambda_k c_k(x) \right]^{\frac{1}{\eta-1}}
\end{equation}
where $Z(\left\{\lambda_i\right\} )\coloneqq\int dx D(x) \left[1+(1{-}\eta)\sum_i \lambda_k c_k(x) \right]^{\frac{1}{\eta-1}} $ is a normalization constant. 
Using a flat default model and constraints on the moments, the above form has a power-law decay for any $\eta<1$.
In the limit $\eta \to 1$, we recover the familiar exponential form of Shannon maximizers.
Spectra with fatter tails appear sharper than ones with the same moments, but which otherwise decay faster.
Therefore, maximizing entropies with lower order $\eta$ is generally expected to produce sharper spectra.
For example, maximaizing Rényi entropy under the constraints of zero mean and variance $\sigma^2$, gives Student's $t$-distributions~\cite{Costa03, Jonson07}
\begin{equation}
	f(x) \propto \left(1+\frac{1{-}\eta}{3\eta{-}1} \frac{x^2}{\sigma^2}\right)^{\frac{1}{\eta-1}}\;,
\end{equation}
which has the full width at half maximum
\begin{equation}
	\operatorname{FWHM} = 2 \sigma \sqrt{\frac{3\eta{-}1}{1{-}\eta} \left(2^{1-\eta}-1\right)}\;.
\end{equation}
As $\eta$ decreases, the tail gets fatter but its FWHM  shrinks, producing a sharper peak.
This readily explains ASM's ability of producing sharper spectra than MaxEnt.
The typical ways of performing ASM correspond to  entropies with orders $\eta = 0$ (sampling only weights) and $\eta = 1/2$ (sampling both weights and positions).
These are lower than the order $\eta = 1$ of MaxEnt's Shannon entropy and lead to peaks with narrower FWHM and fatter tails.
The fat tails also elucidate the observation that ASM on a fixed grid is sensitive to the cutoff/width parameter of a default model, and also why this sensitivity is reduced when the grid positions are sampled~\cite{Ghanem20a, Ghanem20b}.

\begin{figure}
\center
\includegraphics[width=\columnwidth]{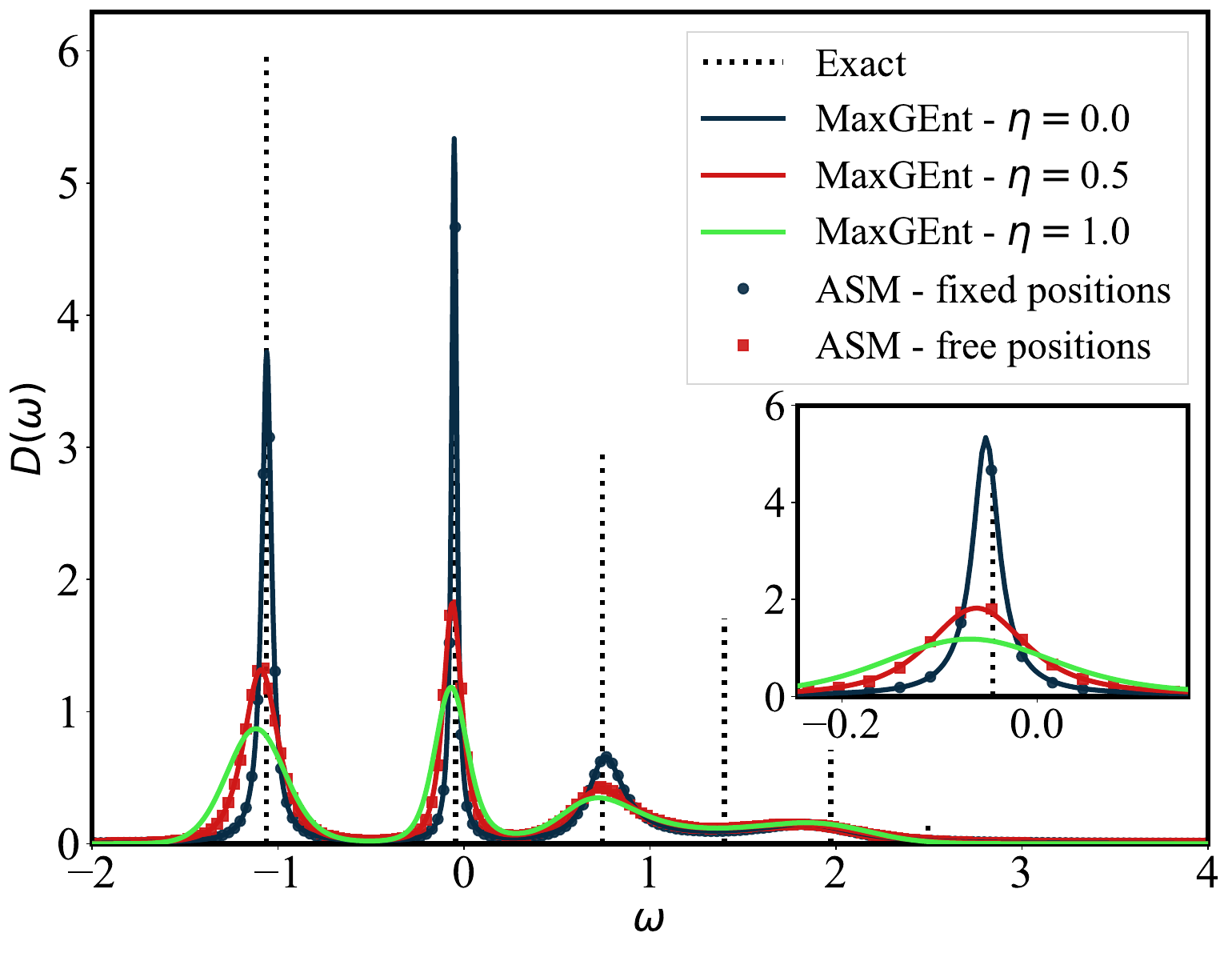}
\caption{\label{fig:spectra} 
    MaxGEnt and ASM results for the analytic continuation of a density of states from the noisy Green function on Matsubara frequencies.
    The exact density is a set of weighted delta peaks which are  plotted here as vertical lines whose heights are propotioanl to their relative weights.
    ASM is performed using $N=256$ delta functions.
    In all methods, a flat default model in the interval $\omega \in [-4, 4[$ is used.
}
\end{figure}

\paragraph{Illustrative Test Case} In practical applications, one does not have access to exact data values, only noisy ones. 
The constraint would then be imposed on the data fit $\chi^2$ which is a quadratic function of the spectrum rather than linear.
Nevertheless, the tendency of lower entropies to produce sharper spectra remains.
As an illustrative example, we look at the analytic continuation problem of the density of states $D(\omega)$ from its Green function $G(z) = 1/(2\pi)\int d\omega D(\omega)/(z{-}w)$ for a hole in the $t$-$J$ model in infinite dimensions~\cite{Strack92} at finite exchange $J/t = 1$.
We evaluated the exact Green function at $8$ Matsubara frequencies $z_j = i (2j{+}1)\pi/8$ and  added relative Gaussian noise with standard deviation $10^{-3}$.
Different entropies are then optimized the under the discrepancy principle constraint (see Fig.~\ref{fig:spectra}).
As expected, using an entropy with lower order produces spectra of sharper peaks and fatter tails, with GK entropy ($\eta=0$) producing the sharpest result.
On the same plot, we also show ASM spectra using both fixed and free positions and under the same discrepancy constraint.
Even using only $N=256$ delta functions, the ASM spectra match the corresponding entropy maximization spectra providing a numerical confirmation of our derivation above.

\paragraph{Summary \& Discussion}
In this letter, we provided a systematic approach for deriving the entropies associated with different ways of sampling in ASM.
This resulted in a modified version of Rényi entropies with the order determined implicitly by the specific way of sampling.
We further showed that the apparently sharp and desirable peaks of ASM can be attributed to the lower order of Rényi entropy associated with typical sampling methods.
These results allow obtaining the same quality of ASM spectra  more  efficiently via maximization of the entropy instead of sampling. 
They also invite a revisit of the unique role of Shannon entropy in MaxEnt.
Besides providing sharper spectra, maximizing Rényi/Tsallis entropies can now be justified by the underlying sampling process.
In particular, sampling the weights with either fixed or free positions is no less justified than sampling the positions alone (which results in Shannon entropy).
This  generalization of MaxEnt gives an additional degree of freedom (the order of the entropy) that allows 
including prior knowledge of the expected profile of the spectral peaks. 

\providecommand{\noopsort}[1]{}\providecommand{\singleletter}[1]{#1}%

\end{document}